\documentstyle[12pt]{article}
\begin{document}
\begin{titlepage}

\begin{center}
{\LARGE  Nambu-Goto string action with \\
Gauss-Bonnet term$^{\star}$}

\vskip 1cm
{\bf J.Karkowski,$^{^{\dag}}$ }
{\bf Z.\'Swierczy\'nski$^{^{\ddag}}$ }
{\bf and P.W\c{e}grzyn$^{^{\dag}}$ } \\
\vskip 1cm

\end{center}

\begin{abstract}

We examine the relativistic Nambu--Goto model with
Gauss--Bonnet boundary term added to the action
integral. The dynamical system is analysed using an invariant
representation of string degrees of freedom
by complex Liouville fields. The solutions of
classical equations of
motion for open strings are described.

\end{abstract}

\vskip 2cm
\begin{tabular}{l}
TPJU--24/94 \\
December 1994
\end{tabular}

\vspace{\fill}

\noindent
\underline{\hspace*{10cm}}

\noindent
$^{\star}$Work supported in part by grant KBN 2 P302 049 05.

\noindent
$ ^{\dag}$Permanent address: Department of Field Theory,
 Institute of Physics,\\
Jagellonian University, Reymonta 4, 30--059 Cracow, Poland.

\noindent
$^{\ddag}$Permanent address: Institute of Physics, Pedagogical
University, \\
Podchor\c{a}\.zych 2,
30--084 Cracow, Poland.

\noindent
\end{titlepage}

Relativistic string theories are often employed to study
the dynamics of one--dimensional objects which may arise
and play a crucial role in some more fundamental
microscopic models. In particular, open strings appear
there as an approximate  description of some realistic
linear objects that spread between some points or sources
(vortices with monopoles at their ends, flux tubes between
quarks etc.). Even though the problem of quantizing
relativistic strings is still cumbrous, it is widely believed
that some kind of useful effective theoretical
string description should exist
for all those various phenomena
 that reveal an appealing 'stringy' character
(e.g. dynamics of linear topological defects in theories
with spontaneous symmetry breaking, fluctuactions of
thin color flux tubes in the confining phase of non--Abelian
gauge theories). The simplest string theory, Nambu--Goto
model, is usually useful to
reproduce some most important qualitative results,
but it fails as we carry the program forward.
There are many directions in which one can modify
the Nambu--Goto model. As far as open strings are concerned,
one might be tempted to embody new features to the system
by considering only interactions between a string and
string ends.

In previous papers \cite{pw,pw2} a suitable formalism was
constructed that described classical dynamics of open strings
with 'interactions at ends'. In light of these works,
we consider here a particular example of extended Nambu--Goto model,
when Gauss--Bonnet term is added to the action. This model
has been examined earlier in other works [3--7].
Zheltukhin \cite{cc1} considered the system embedded in
three--dimensional spacetime and found the correspondence
between classical string solutions and solutions of
the real Liouville equation with Dirichlet--type boundary
conditions. However, only  static solutions have been
found and discussed. Barbashov and collaborators [4--6] worked out
the classical system of equations eliminating gauge freedom
(choice of parametrization) and derived some particular
type of solutions.
In the presented paper, we re-examine the model and find all
types of classical solutions. The most interesting result is
that only planar string configurations can be extrema of
the classical action.

The classical string model discussed by us is  defined by the
following action integral:
\begin{equation}
\label{act}
S = -\gamma A - \frac{\alpha}{2}S_{GB} \ ,
\end{equation}
where $\gamma$ stands for string tension,
$A$ denotes world sheet area,
 $\alpha$ is a dimensionless parameter.
Parameters $\gamma$ and $\alpha$  are arbitrary positive real constants.
 $S_{GB}$ is a pseudoeuclidean version of Gauss--Bonnet term, being
the total integral of internal world sheet curvature.
The string action (\ref{act}) shows no difference with
the two--dimensional gravity model, a cosmological (Nambu--Goto)
term is supplemented by Einstein  (Gauss--Bonnet) term. As
Einstein Lagrangian is a total derivative in two dimensions,
this action term acts only at boundaries, modifying
the sector of open strings.

Classical string equations of motion derived from (\ref{act}) are
Nambu--Goto equations together with some 'edge conditions',
being in fact dynamical equations involving third order
time derivatives.
As usual in the case of minimal surfaces, their general solution can
be represented as the combination of left-- and right--moving parts,
\begin{equation}
\label{sss}
X_{\mu}(\tau,\sigma) = X_{L\mu}(\tau+\sigma) + X_{R\mu}(\tau-\sigma),
\end{equation}
where the world sheet parametrization has been  defined by the
following conditions \cite{bn}:
\begin{eqnarray}
\label{param}
(\dot{X} \pm X')^2 & = & 0, \nonumber \\
(\ddot{X} \pm \dot{X}')^2 & = &  - q^2.
\end{eqnarray}
Bulk equations of motion get linearized due to the orthonormal gauge.
As this gauge still allows for
conformal changes of parametrization,
so that this residual gauge freedom has been saturated by the
latter pair of equations in (\ref{param}).
The parameter $q$ has dimension of mass and can take any positive
real value. This is an integral of motion, which fixes the scale
for considered solutions.

Up to transformations of Poincare group, minimal surfaces $X_\mu$
para\-me\-trized according to (\ref{param}) correspond to solutions
$\Phi$ of the complex Liouville equation \cite{pw,pw2}:
\begin{equation}
\label{liouv}
\ddot{\Phi} - \Phi'' = 2q^{2}e^{\Phi}.
\end{equation}
This correspondence can be presented in the following way:
\begin{eqnarray}
\label{sol}
e^{\Phi} & = & -\frac{4}{q^2}
\frac{f_{L}'(\tau+\sigma)f_{R}'(\tau-\sigma)}
     {[f_{L}(\tau+\sigma)-f_{R}(\tau-\sigma)]^2} \ , \nonumber \\
\dot{X}^{\mu}_{L,R} & = &
\frac{q}{4|f_{L,R}'|}(1+|f_{L,R}|^2,~2\mbox{Re}\,f_{L,R},
{}~2\mbox{Im}\,f_{L,R},
                      ~1-|f_{L,R}|^2) \ ,
\end{eqnarray}
where $f_L$ and $f_R$ are arbitrary complex functions.
Note that any simultaneous modular transformation of $f_{L}$ and
$f_{R}$ induces Lorentz transformation of $X_{\mu}$ while
Liouville field $\Phi$ remains invariant.
For later discussions, it is helpful to invert relations (\ref{sol}):
\begin{eqnarray}
\label{invsol}
S(f_{L,R}) = \partial^{2}_{\pm} \Phi - \frac{1}{2} (\partial_{\pm}
            \Phi)^{2} \ , \nonumber \\
f_{L,R} = \frac{\dot{X}_{L,R}^{1} + i \dot{X}_{L,R}^{2}}{
           \dot{X}_{L,R}^{0} +  \dot{X}_{L,R}^{3}} \ ,
\end{eqnarray}
where $\partial_{\pm} = \frac{1}{2} (\partial_{0} \pm \partial_{1})$,
and $S(f)$ stands for Schwartzian derivative:

$$S(f) = \frac{f'''}{f'} - \frac{1}{2} \left(
\frac{f''}{f'} \right)^2 \ .$$

All classical solutions of the Nambu--Goto string model with
Gauss--Bonnet boundary term added, represented as above (\ref{sol}),
are subject to the following conditions imposed at string
ends \cite{pw2,hw}:
\begin{eqnarray}
\label{bc1}
e^{\Phi} & = & -\frac{1}{q}\sqrt{\frac{\gamma}{\alpha}}
              \equiv - \frac{C^2}{q^2}
                  \ , \\
\label{bc2}
\mbox{Im} \, \Phi' & = & 0 ~~~ \mbox{for}~~ \sigma = 0, \ \pi \ .
\end{eqnarray}

It is helpful for later analysis to discuss possible singularities
of Liouville fields. Since ${\rm Im} \Phi$ is
some angle variable, the singularities refer only to
${\rm Re} \Phi$.
Singular points
can be taken into account only if ${\rm Re} \Phi > 0$
in their vicinities
\cite{pw2}.
It implies that at any world sheet point we have \cite{pw2}:
\begin{eqnarray}
\mbox{a) } & \mbox{ derivatives} \
f'_{L} \  \mbox{and}  \ f'_{R} \
\mbox{cannot take zero values;}
\nonumber \\
\mbox{b) } & \mbox{ ratios} \ f_{L}^{2}/f'_{L}\  \mbox{and}
 \ f_{R}^{2}/f'_{R}
\ \mbox{are finite}
\label{req}
\end{eqnarray}

The singularities of the
Liouville field with ${\rm Re} \Phi > 0$ correspond to
singularities of the induced metric: $g=0$.

Let us now start to study boundary conditions (\ref{bc1}, \ref{bc2}).
First, we note that Liouville field $\Phi$ and its first and second
derivatives are real at boundary points $\sigma=0, \pi$.
Thus, it follows from relations (\ref{invsol}) that Schwartzian
derivatives of $f_{L}$ and $f_{R}$ are real for any values
of their arguments. Henceforth, one can always find a modular
transformation that turns either $f_{L}$ or $f_{R}$ into unimodular
functions. Let us specify the following choice:
\begin{equation}
f'_{L} = e^{i F_{L}} \ \ , \ \ \ f'_{R} = \frac{a e^{i F_{R}} +
b}{c e^{i F_{R}} + d} \ ,
\end{equation}
where $F_{L}$ and $F_{R}$ are some real functions, and complex
coefficients of the modular transformation satisfy: $ad - bc =1$.
Furthermore, taking into account imaginary part of Eq.(\ref{bc1}) and
Eq.(\ref{bc2}) one can easily obtain that
\begin{equation}
a = \pm \bar{d} \ \ , \ \ \ b = \pm \bar{c} \ .
\end{equation}
This implies that the function $f_{R}$ must be unimodular as well.
After some simple redefinition, we have now
\begin{equation}
\label{fF}
f_{L} = e^{i F_{L}} \  \mbox{ and } \     f_{R} = e^{i F_{R}}
\end{equation}
and
\begin{eqnarray}
\label{df1}
e^{\Phi} & = & -\frac{1}{q^2}\frac{F_L'F_R'}{\sin^2\frac{F_L -
                F_R}{2}}\ , \\
\label{df2}
\dot{X}^{\mu}_{L,R} & = & \frac{q}{2|F_{L,R}'|}(1,~\cos F_{L,R},
                          ~\sin F_{L,R},~0)\ .
\end{eqnarray}

As a consequence of the boundary conditions, the imaginary part of
Liouville field $\Phi$ is constant and equal $\pi$ at any point
of the world sheet. Thus, a surprising conclusion has been drawn
from the above analysis of boundary conditions (\ref{bc1},\ref{bc2})
-- any particular classical solution of the equations of motion
describes a string which time evolution entirely takes place
in some fixed plane. In other words, dynamics of extrinsic geometry of
the world sheet (described by ${\rm Im} \Phi$) is frozen. The world
sheet is immersed in some three--dimensional
plain hyper--surface. A unit vector normal to this hyper--surface
is an integral of motion and can be used to specify different
solutions.

Next, from boundary conditions (\ref{bc1})
and from requirements (\ref{req}) we obtain that
\begin{equation}
\label{nons}
F'_{L} F'_{R} > 0 \ ,
\end{equation}
at any point of the world sheet. It means that both real functions
$F_{L}$ and $F_{R}$ are either increasing or decreasing.

If we have already shown that only planar solutions can exist,
let us attempt to solve
boundary problem (\ref{bc1},\ref{bc2}) for the complex Liouville
equation
completely. We will show that all solutions can be divided into
three groups:

A. periodic solutions:
$$\ f_{L}(\tau)=f_{L}(\tau+2\pi) \ \mbox{and}
                       \ f_{R}(\tau)=f_{R}(\tau+2\pi) \ ;$$

B. 'furled' strings:
$$ \ f_{L} (\tau) = \lambda \tau \ \mbox{and}
                   \ f_{R} (\tau) = \lambda (\tau + \pi) ;$$

C.  non--periodic solutions:
$$\ f_{L}(\tau) \ \neq f_{L}(\tau+2\pi) \ \mbox{and}
                 \ f'_{L}(\tau) \neq  f'_{L}(\tau+2\pi) \ ,$$
or:
$$\ f_{R}(\tau) \ \neq f_{R}(\tau+2\pi) \ \mbox{and}
                 \ f'_{R}(\tau) \neq  f'_{R}(\tau+2\pi) \ .$$

It is important to note that we consider only the period $2\pi$,
where $\pi$ is distinguished here as it is
the length of $\sigma$--interval. Therefore,
periodic solutions with other periods, if they exist,
do not belong to the class A.

{\em CASE A:}

In the case A, solutions have the following general form:
\begin{eqnarray}
F'_{L} = \frac{1}{2} \left[ h' \pm
 \sqrt{h'^2+4 C^2 \sin^2{(h/2)}}
           \right] \ , \nonumber \\
F'_{R} = \frac{1}{2} \left[ - h' \pm
 \sqrt{h'^2+4 C^2 \sin^2{(h/2)}}
           \right] \ ,
\label{solA}
\end{eqnarray}
where constant $C^2$ has been defined in (\ref{bc1}) and
$h$ is some periodic (with period $2\pi$)
function. Its arbitrariness is restricted by the
following constraints:
\begin{eqnarray}
h(\tau+2\pi) = h(\tau) \ , \nonumber \\
h(\tau) \neq 2k\pi \ \mbox{for any} \ \tau \ \mbox{and} \
\mbox{integer} \ k \ , \nonumber \\
\int_{0}^{2\pi} d\tau \sqrt{h'^2 + 4 C^2 \sin^2{(h/2)}} =
2 N \ ,
\label{reqca}
\end{eqnarray}
where $N$ is some natural number. The conditions (\ref{reqca})
follow from periodicity (case A) and non--singularity (\ref{nons})
requirements.
 One can concince oneself
that there exist infinitely many functions $h$ obeying the above
requirements, so that there are infinitely many examples
of periodic solutions.

As $h$ in (\ref{solA}) we can take an arbitrary periodic function
which is a mapping on
any finite interval that does not include points
$2k\pi$. For any given $N$, we can always match a contant
of motion $C$
according to (\ref{reqca}). Finally, periodicity requirements
stated in type A are assured if we adjust integration constants
while solving (\ref{solA}) in such a way that
\begin{equation}
F_{L} - F_{R} = h \ .
\end{equation}

{\em CASE B:}

If some non-periodic function either $f_{L}$ or $f_{R}$ has a periodic
first derivative, then one can derive easily from boundary
conditions (\ref{bc1},\ref{bc2}) that $f_{L}''=f_{R}''=0$.
Therefore, we obtain the solution B, which represent a folded straight
string propagating with some longitudinal velocity.
The constant $q$ is given by

$$ q = \frac{4}{\pi^2} \sqrt{\frac{\alpha}{\gamma}} \ .$$

{\em CASE C:}

In the case $C$, boundary conditions (\ref{bc1}) enable us
to find the explicite relation between left- and right-movers:
\begin{eqnarray}
F_{R}(\tau) = F_{L}(\tau+2\pi) + \nonumber \\
\mbox{arc ctg} \Big[ \pm
\sin^{-1}{(\Delta F_{L})} - \mbox{ctg}(\Delta F_{L}) \Big] \ ,
\end{eqnarray}
where

$$\Delta F_{L} \equiv \frac{F_L(\tau+2\pi)-F_L(\tau)}{2} \ .$$
The function $F_{L}$ itself is subject to the following equation:
\begin{eqnarray}
\frac{F''_L(\tau+2\pi)}{F'_L(\tau+2\pi)} -
\frac{F''_L(\tau)}{F'_L(\tau)} -
[F'_L(\tau+2\pi) + F'_L(\tau)] \mbox{ctg}(\Delta F_L)
\nonumber \\
= \pm \frac{C^2 - 2 G^2_L}{G_L} \ ,
\label{solC}
\end{eqnarray}
where

$$G_L \equiv \frac{\sqrt{F'_L(\tau+2\pi) F'_L(\tau)}}{
\sin{(\Delta F_L)}} \ .$$

First, let us discuss how many solutions of type $C$ there exist.
Let us freely define the function $F_{L}$ in the interval
$[0,2\pi]$ ensuring only that the values of the
function and its first and second
derivatives  at points $0$ and $2\pi$ satisfy (\ref{solC}).
Now, we can extend the definition of $F_{L}$
to the interval $[2\pi,4\pi]$ (or to the interval $[-2\pi,0]$)
by solving differential equation (\ref{solC}) for $F_{L}(\tau)$
with given retarded $F_{L}(\tau-2\pi)$  (or advanced
$F_{L}(\tau+2\pi)$) values taken from the interval $[0,2\pi]$.
Following this procedure,
starting from arbitrary  ${\cal C}$ function
$F_{L}$ defined in the interval
$[0,2\pi]$ which values at end points
satisfy algebraic equations (\ref{solC}),
we can construct a
${\cal C}^2$ solution to differential equation (\ref{solC}).
The construction described here (see also \cite{r1,r2})
provides also a basis
of an algorithm to generate numerical solutions.

The above construction can be applied unless
the function $e^{iF_L}$
is periodic. If it happens, we can rearrange the
construction using the other function $F_R$.
However, in case either $F_L$ or $F_R$ is periodic
we can use a simpler method to derive solutions.
We are looking for the function $h=F_L-F_R$ which can
be derived either from the equation
\begin{equation}
(F'_{L})^2 - F'_L h' = C^2 \sin^2{ \frac{h}{2} } \ ,
\label{kark}
\end{equation}
with a given periodic function $e^{i F_L}$, or from the equation
\begin{equation}
(F'_R)^2 + F'_R h' = C^2 \sin^2{\frac{h}{2}} \ ,
\end{equation}
with a given periodic function $e^{i F_R}$.

Let us give some examples of solutions.
The rotating rigid rod solution \cite{hw} is given by
\begin{equation}
\label{eq22}
F^{(0)}_L(\tau) = \pm \lambda(\tau - \frac{\pi}{2}) \ , ~~~
F^{(0)}_R(\tau) = \pm \lambda(\tau + \frac{\pi}{2}) + \pi \ ,
\end{equation}
where frequency $\lambda$ satisfies

$$ \lambda^2 = q \sqrt{\frac{\gamma}{\alpha}}
\cos^2{\frac{\lambda \pi}{2}} \ .$$
The plus and minus signs correspond to clockwise and
anti--clockwise revolutions respectively.

The rotating rigid rod solution is a static (solitonic) solution
of the Liouville equation (\ref{liouv}). We perturbed this solution
and solving numerically the equation (\ref{solC}) (according to
described algorithm) we obtained a solution depicted in Fig.1
($F'_L$ is represented here).
This exact numerical solution can be compared with the approximate
general solution derived in \cite{hw} for perturbations around
the static solution:
\begin{equation}
\label{eq23}
F_{L,R} = \frac{\pi}{2} \mp \frac{\pi}{2}
\pm\sum_{n/1}^{\infty}D_n\sin\left[\omega_n(\tau
\mp \frac{\pi}{2}) + \varphi_n \pm \frac{n\pi}{2}\right],
\end{equation}
where plus and minus signs correspond to left-- and right--movers
respectively, and $D_n$, $\varphi_n$ are arbitrary real constants.
The eigenfrequencies $\omega_n$ can be calculated from
the transcendental equation:

$$ \omega_n \mbox{tg}{ \frac{\pi(\omega_n + n)}{2} } =
\lambda \mbox{tg}{ \frac{\pi\lambda}{2} } \ .$$

An approximate solution obtained from the above series,
which corresponds to the same initial data as the numerical one,
is almost exactly the same as that plotted in Fig.1.

The special case associated with the equation (\ref{kark}) can
be illustrated with the following exact solution:
\begin{eqnarray}
F_L & = & N \tau \ , \nonumber \\
F_R & = & N \tau + 2 \mbox{arc tg}(e^{N\tau}) - \frac{3\pi}{2} \ ,
\end{eqnarray}
where $N$ is integer and $C^2 = 2 N^2$.

This work was supported in part by the KBN under grant 2 P302 049 05.

{\bf \large Figure Caption}

Fig.1. A numerical solution to the equation (20). The first derivative
of $F_L$ is plotted.

\end{document}